\documentclass{article}

\usepackage{arxiv}

\usepackage[utf8]{inputenc} 
\usepackage[T1]{fontenc}    
\usepackage{hyperref}       
\usepackage{url}            
\usepackage{booktabs}       
\usepackage{amsfonts}       
\usepackage{nicefrac}       
\usepackage{microtype}      
\usepackage{lipsum}
\usepackage{hyperref} 
\usepackage{blindtext}
\usepackage{enumitem,amssymb}
\usepackage{pifont}
\usepackage{mathrsfs}
\usepackage{color}
\usepackage{multirow}

\usepackage{listings}
\usepackage{float} 
\usepackage{subfigure} 
\usepackage{graphicx}

\definecolor{gray}{rgb}{0.5,0.5,0.5}

\title{Using GGNN To Recommend Log Statement Level}

\author{
  Mingzhe Li\\
  Department of Electronic Engineering\\
  Tsinghua University\\
  Beijing, China \\
  \texttt{li-mz16@mails.tsinghua.edu.cn} \\
   \And
 Jianrui Pei \\
  Department of Electronic Engineering\\
  Tsinghua University\\
  Beijing, China \\
  \texttt{pjr16@mails.tsinghua.edu.cn} \\
   \AND
 Jin He \\
  VMware \\
  Beijing, China \\
  \texttt{hejin@vmware.com} \\
   \And
  Kevin Song \\
  VMware \\
  Beijing, China \\
  \texttt{skevin@vmware.com} \\
   \And
  Yongfeng Huang \\
  Department of Electronic Engineering \\
  Beijing, China \\
  \texttt{yfhuang@tsinghua.edu.cn} \\
   \And
  Frank Che \\
  VMware \\
  Beijing, China \\
  \texttt{fche@vmware.com} \\
   \And
  Chitai Wang \\
  VMware \\
  Beijing, China \\
  \texttt{wangch@vmware.com} \\
}

\begin{document}
\maketitle

\begin{abstract}
In software engineering, log statement is an important part because programmers can't access to users' program and they can only rely on log message to find the root of bugs. The mechanism of "log level" allows developers and users to specify the appropriate amount of logs to print during the execution of the software.\cite{hengloglevel} And 26\% of the log statement modification is to modify the level.\cite{clpioss} We tried to use ML method to predict the suitable level of log statement. The specific model is GGNN(gated graph neural network) \cite{ggnn} and we have drawn lessons from Microsoft's research.\cite{AllamanisProgramGraph} In this work, we apply Graph Neural Networks to predict the usage of log statement level of some open source java projects from github. Given the good performance of GGNN in this task, we are confident that GGNN is an excellent choice for processing source code. We envision this model can play an important role in applying AI/ML technique for Software Development Life Cycle more broadly.
\end{abstract}


\section{Introduction}
When a system breaks down, developers usually can't access to it because the program runs on user's computer. The only thing on which they can rely is the log statement. A log statement can report where the error takes place and the value of some critical variables so that developers can give a diagnosis. Ideally, coders can put log statements after every non-log statement. Theoretically speaking, doing like this eliminates ambiguity to the greatest extent. Unfortunately, it will introduce unbearable performance overhead. So we have to consider where to place a log statement is a best choice. Currently, there are some tools which can help programmers to make a decision whether to write a log statement. For example, Log20 can automates log statement placement without any domain knowledge. \cite{log20} But it's not enough to just locate log statements. In engineering, an experienced coder is often aware of which lines are turning points and that he should place a log statement there. The question is how to make a trade off between system overhead and program running information. Using log verbosity level is a solution. Common log libraries like Log4j and SLF4J support six log levels, including \textit{trace, debug, info, warn, error} and \textit{fatal}. Users can control the verbosity level of log statements to be printed out during execution. \cite{hengloglevel} By doing this, debuggers can adjust the amount of log information that they receive dynamically. However, assigning the right verbosity level requires sufficient experience. Developers spend significant efforts on adjusting the verbosity level of log messages, accounting for 26\% of all the log improvements. Majority (72\%) of them reflect the changes in developers' judgment about the criticality of an error event.\cite{clpioss} We build a machine learning model based on GGNN to help developers to choose the right verbosity level.

When it comes to using machine learning to solve problems in software engineering, most of the state-of-art projects are concentrated on natural language processing (NLP) method. Indeed, source code and natural language are similar in many ways, such as they are both made up by limited characters and symbols and are both sequential. But we still can't treat them equally. 
As the size of vocabulary of the source code is beyond imagination while the same of natural language is limited, we can find that most of the words in the code are created temporarily, and they are not used a lot. That means it is tough to learn the semantic information of these words.
For example, the number of total tokens of WMT-14 English dataset is 304,000,000 and the size of vocabulary of it is 160,000, while there are 12,095,591 tokens in Eclipse 3.5.2 and its vocabulary size is 291,671. \cite{deepautocoder} All of these leads to difficulties in applying NLP to source code. The paradox is that some simple non-deep-learning method even outperforms complicated deep learning model in this field. \cite{isDLgood} We assume the reason why some deep learning models don't perform well is that they can't take structural information into account and most identifiers are unknown for them so that they can't extract enough semantic information. Maybe we should change our perspective. Code is more like being organized as a graph. Whenever there is a function call, a variable definition or an operational expressions, an edge will be linked. So graph neural networks (GNN) is also a possibility because it can describe the structure information of the program by setting up some edges. GGNN is a kind of GNN. It realizes the information transfer between nodes through GRU. We use GGNN to learn the representation vector of a log statement and get a log level prediction by analysing the vector with a MLP.

We make the following contributions: 1)We build a model based on GGNN to predict log level. It can recommend level of a log statement according to the program subgraph whose center is the log statement. 2)We evaluate this model on 29 open-source java projects from Apache on Github and compare it with random model , LSTM model and Heng Li's model. \cite{hengloglevel} The result shows that this model outperforms them.

\section{Related Work}
Our work is based on prior researches on log sentences and log levels. Zhao et al.(2017)\cite{Zhao2017Log20FA} presented log20 to determine a near optimal placement of log printing
statements. Yuan et al.(2010)\cite{Yuan2010SherLogED} analyzes source code by leveraging information provided by run-time logs to infer what must or may have happened during the failed production run. Zhu et al.(2015)\cite{Zhu2015LearningTL} proposed LogAdvisor that extracts contextual features and helps developers to determine where to log.
\\This model is constructed on recent works of representing source code in machine learning models. Alon et al.(2018a)\cite{Alon2018code2seqGS}, Alon et al.(2018b)\cite{Alon2018Code2vecLD} represent code snippets as a set of compositional paths over its abstract syntax tree (AST) and use LSTM to compress the paths into fixed-length vector.  Allamanis et al.(2014)\cite{Allamanis2014LearningNC} introduce NATURALIZE to learn local context and solve coding convention inference problem. Vasic et al.(2019)\cite{Vasic2019NeuralPR} construct multi-head pointer network to perform joint prediction of both the location and the repair for variable misuse bugs. 
\\Our work is closely related to the research of Heng Li et al.(2017)\cite{hengloglevel} who analyze the key features of log level through case study and represented the features with arrays. However, they manually extract the features and do not use the semantic information of the code.
\\Graph neural network is used to extract the syntactic and semantic features of source code in this model. Gori et al.(2005)\cite{Gori2005ANM} proposed GNN to cope with graphical data structures and Li et al.(2016)\cite{Li2016GatedGS} built GGNN for non-sequential outputs. Allamanis et al.(2017)\cite{AllamanisProgramGraph} adapted GGNN to program graphs and achieved excellent results in variable misuse task. Brockschmidt et al.(2018)\cite{Brockschmidt2018GenerativeCM} extended the model of Allamanis et al.(2017)\cite{AllamanisProgramGraph} to code generation and achieved in hole completion task.

\section{Log Verbosity Level Prediction Task}

Whatever programming language you use, printing log information is necessary. The main purposes of printing log information are as follows: 1.We can find bugs in programs through it. 2.We can monitor the status of the system and prevent problems from happening through it. 3.We can ferret out unsafe, unauthorized operations to guard against attacks from others. Although there are so many benefits, the best log library still have impact on performance. To minimize the overhead, one solution is to assign different levels to different log statements and print them by level in appropriate situations. Generally speaking, the log level can be divided into six categories: trace, debug, info, warn, error and fatal. Trace-level statements are the most detailed and mainly used in the development phase. Once being deployed to production environments, they should be deleted. Debug-level statements are similar to trace-level statements. The difference is that the former lasts longer. Info-level statements are used to print some normal information. Warn-level statements mean that there are some potential problems, but the program can go on. Error-level statements appear when some very serious problems occur. These problems will cause the current program to fail to execute properly. But if an exception handling mechanism exists, the system will not crash. Finally, fatal-level statements indicate that the worst case occurs, that is, the system crashes. Some of these levels are easily confused, such as trace and debug, warn and error. Choosing the right log level consumes a lot of energy for programmers, 26\% of all the log improvements are targeted at adjusting the log level. \cite{clpioss}

There is a correlation between the log statements and their program contexts, because they need to appear in the event of problems and print some related variables if necessary. If they don't appear in particular locations, the values of these variables may change and become meaningless. So it's possible to judge a log statement level from its context. Our goal is to build a model for predicting log levels based on program context.

\section{Model}

This model is based on GGNN\cite{ggnn}, and the definitions of node and edge are inherited from a related Microsofts' research. \cite{AllamanisProgramGraph}Considering the characteristics of this task, we use MLP to handle the output vector of GNN.

\subsection{Graph Neural Networks}

The first part of this model is GNN and we summarize it here. A graph $\mathscr{G}$ = ($\mathscr{N}$, $\mathscr{E}$, $\mathscr{X}$) has three parts: a set of nodes, a list of edges and initial representations of nodes, and there can be several kinds of edges. Edges are pairs $e = (v_1, v_2) 
\in \mathscr{N} \times \mathscr{N}$. The initial representations of nodes are generally a list of real vector ${x}$. The purpose of GNN is to learn node vector which is donated by $h$. Vector $x_n$ and vector $h_n$ have different meanings, the former is our prior knowledge about a node while the latter is the output of GNN with graph message.

Generally, GNN works like this: firstly there are several iterations of propagation that compute every node representation vector $h$; secondly there is a output function $g(h, x)$ that calculate the output that we want to get. In order to get a prediction of log level, every node from which we want to get output is a log statement, and their initial representations are the same, so the output function can be simplified to $g(h)$. The parameters of output function $g(h)$ can be learned and we use MLP as function $g(h)$.

In this section, we will elaborate on the propagation step. At the first, we assign the value of every initial node vector $x_n$ to $v_{n0}$. Generally, the sizes of $x$ and $v$ are the same. However we can enlarge the size of vector $v$ by padding zeros at the end of it. In each iteration process, every node sends message to its neighbors. The content of messages is determined by both node vectors and the type of edges:

$$
m_{kn} = f_k(v_n)
$$

$m_{kn}$ means the message which node $n$ sends through edge of type $k$. And the parameters of function $f_k$ can be learned. Referring to the method in \textbf{LEARNING TO REPRESENT PROGRAMS WITH GRAPHS}\cite{AllamanisProgramGraph}, we set this function to be linear. For those nodes which have plural neighbors, a process of aggregating messages is necessary before updating. There are two simple methods to aggregate messages: maximizing or averaging. In each iteration, every node vector $v$ is updated at the same time. The update operation is as follows: once every node gets its aggregated message $\tilde{m}$, the node vector of next step is calculated by:

$$
h^{'}_n = GRU(h_n, \tilde{m})
$$

GRU is the recurrent cell function of gated recurrent unit.\cite{gru} After a certain number of iterations, the final node vector contains not only the information of the node itself, but also the information of a subgraph whose center is the node.

\subsection{Program Graphs}

In order to apply GGNN to processing code, we have to convert source code to graph. In fact, source code is inherently graphical. We use the definitions of nodes and edges in \textbf{LEARNING TO REPRESENT PROGRAMS WITH GRAPHS}\cite{AllamanisProgramGraph} and we will elaborate on them in this section. 

\large node type:

\normalsize There are 7 kinds of node, and their definition is in Table 1.

\begin{table}[h!]
\Large  
\caption{node type}  
\fontsize{12}{15}\selectfont 
\begin{center}  
\begin{tabular}{|p{100pt}|p{280pt}|}  
\hline  
TOKEN & Corresponding to keywords and punctuations\cr\hline 
IDENTIFIER TOKEN & Corresponding to identifiers created by the programmer\cr\hline
COMMENT LINE & Corresponding to comments in source code\cr\hline
AST ELEMENT \& FAKE AST & Corresponding to some nodes used to describe AST structure\cr\hline
SYMBOL TYP & Corresponding to some nodes indicating that some other node are type identifiers\cr\hline
TYPE & Corresponding to variable types\cr\hline
\end{tabular}  
\end{center}  
\end{table}  

\large edge type:

\normalsize There are 14 kinds of edge, and I will exemplify their meaning.
When there is an AST CHILD edge between two nodes, these two nodes are adjacent in AST. When there is a NEXT TOKEN edge between two nodes, these two nodes are adjacent in source file. A LAST WRITE edge links an IDENTIFIER TOKEN node to where it last appears in the source file. A LAST USE edge is similar to a LAST WRITE edge but it only links two IDENTIFIER TOKEN nodes when the former is used in that line, not defined. For example:  

\begin{lstlisting}
int i;
i = 9;
\end{lstlisting}
in the code block above, there is only LAST WRITE edge between two "i" nodes while in the code block below, there are both LAST WRITE edge and LAST USE edge.
\begin{lstlisting}
i = 0;
x = i + j;
\end{lstlisting}

A COMPUTED FROM edge links a node to another node from which it is computed. Here is an instance in which there is a COMPUTED FROM edge between "x" and "y":

\begin{lstlisting}
x = y + 5;
\end{lstlisting}

A RETURNS TO edge links "return" token to the method declaration. A FORMAL ARG NAME edge links arguments in the method calls to the formal parameters that they are matched to. \cite{dasd} Here is an instance in which there is a COMPUTED FROM edge between "param" and "x":

\begin{lstlisting}
private static int myfunc(int param) {
...
}

myfunc(x);
\end{lstlisting}

GUARDED BY and GUARDED BY NEGATION edges are used to connect every token corresponding to a variable to enclosing guard expressions. Here is an instance in which there is a GUARDED BY edge between "x" and "x>y" and a GUARDED BY NEGATION edge between "y" and "x>y", "x>y" is an AST node:

\begin{lstlisting}
if (x>y){
    ...x...
}
else{
    ...y...
}
\end{lstlisting}

LAST LEXICAL USE edges link all variables with the same name. LAST LEXICAL USE is different from LAST WRITE because two variables can occupy different memory spaces while they share the same name. This phenomenon is particularly prone to occur when there are local variables. An ASSOCIATED TOKEN edge links a TOKEN or IDENTIFIER TOKEN node to its corresponding AST node. A HAS TYPE edge links an object to its corresponding TYPE node. An ASSOCIATED SYMBOL edge links a type identifier to its corresponding SYMBOL TYP node.

Figure 1 shows an example of program graph.

\begin{figure}[H] 
\centering 
\includegraphics[width=\textwidth]{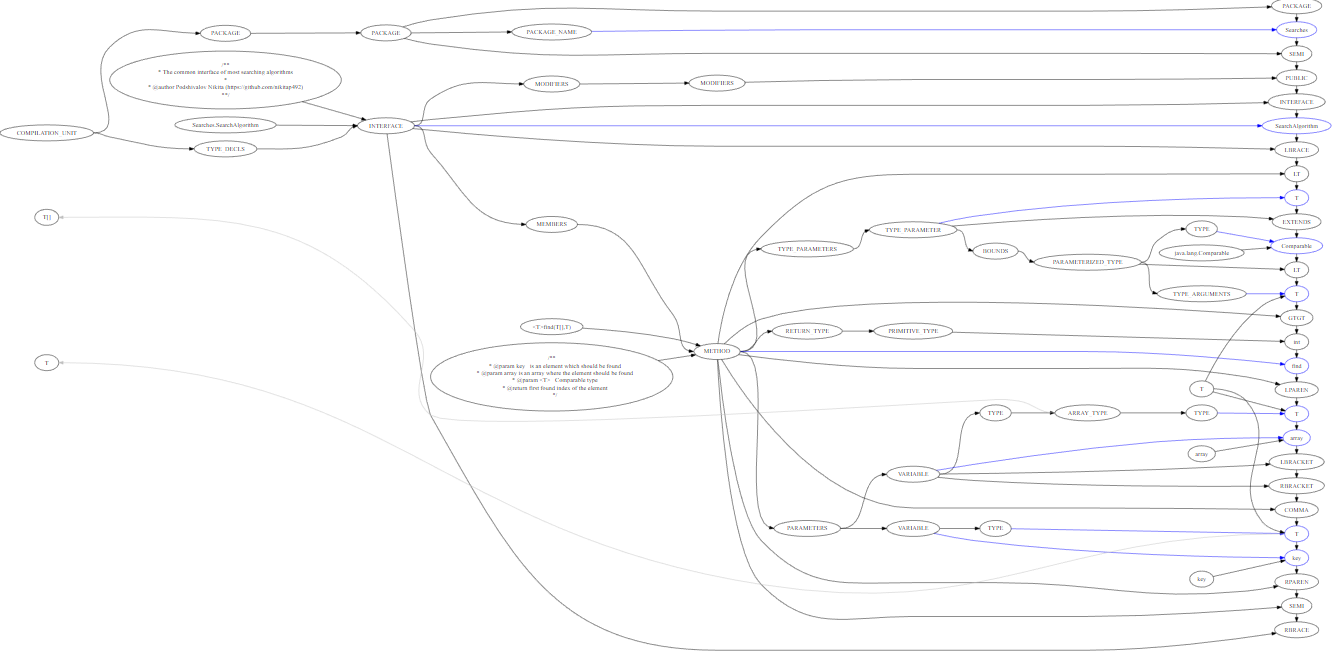} 
\caption{Program Graph} 
\end{figure}

\subsection{MLP}

MLP is the simplest neural network. The output of GGNN is supposed to be a real vector and it is suitable for processing with MLP. Now, our task is to divide the vector into six categories. So we construct a neural network consisting of several full-connected layers to extract the features of vectors automatically. The output of it is a vector of length 6. The value of the n-th component of this vector represents the possibility that the log statement is of level n.

\vspace{12pt}

Figure 2 shows the overall framework of the model.

\begin{figure*} 
\centering 
\includegraphics[width=0.8\textwidth]{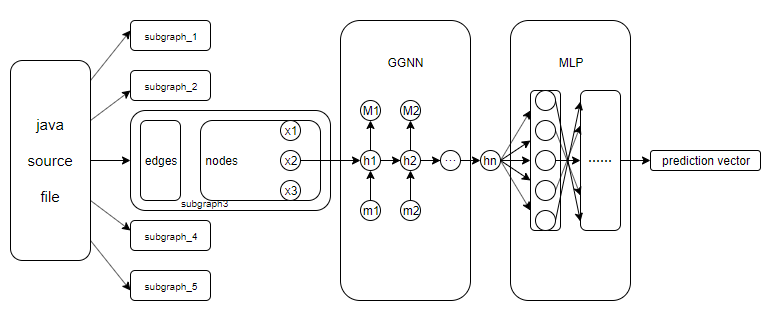} 
\caption{Framework} 
\end{figure*}

\section{Implementation Details}

\subsection{preprocessing method}

The goal of preprocessing is to convert java source code to a number of labeled log statement-centric subgraph. We use an open-source java tool, features-javac, to accomplish the first step.\footnote{https://github.com/acr31/features-javac} It not only converts source code to graph but also connects all defined edges. The graph are saved as proto format and then parsed with Google's protobuf library. \footnote{https://github.com/protocolbuffers/protobuf} 

Sometimes log statements are hard to figure out. Different systems usually have log statements in different formats. In order to locate log statement and ascertain its verbosity level, we have to ensure that the program uses common log library. There are several common log library, e.g. Log4j \footnote{http://logging.apache.org/log4j/2.x} and slf4j \footnote{http://www.slf4j.org}. Using these libraries, log statements are like the following:

\begin{lstlisting}
if (args!= null){
    int argNb = 0;
            
    for (String arg:args){
        LOG.info("Args[{}]:{}", 
        argNb, arg);
        argNb++;
    }
}
\end{lstlisting}

It's easy to determine whether a line of code is a log statement by identifying whether there is a string "log" and a verbosity level keyword in it. A simple script can accomplish this task.

\subsection{subgraph extraction}

The purpose of this model is to predict the verbosity level of a log statement according to a graph containing it. But we have to get rid of the log statement' own information. In order to ensure that the processed program can still be compiled so that graphs can be generated, we simply replace the log statement with a semicon. For example, the code block above will become like this after processing:

\begin{lstlisting}
if (args!= null){
    int argNb = 0;
            
    for (String arg:args){
        ;
        argNb++;
    }
}
\end{lstlisting}

Considering that the number of steps of propagation is limited, a subgraph already meets our needs. Then what we should do is to generate a subgraph whose center is the semicon. Specific approach is to add all points within several steps from the semicon to the node set of the subgraph and then add all edges whose starting node and ending node are in the node set to the edge set of the subgraph. In fact, selection of subgraph is pretty flexible. More steps will introduce more information and the model may learn more complex features while the problem of data sparsity will arise. Less steps helps to learn some direct or superficial features. Selecting nodes within m steps and outside n steps can help to exclude some unessential features.

\subsection{token embedding}

A universal problem in applying NLP or other ML methods to code is countless kinds of tokens. An important reason of this problem is that programmers prefer to combine several words into an identifier token. Generally, they do this with UnderScoreCase or camelCase. Once we split every identifier into several parts, the vocabulary is reduced to an acceptable level. Inspired by Deep-AutoCoder\cite{deepautocoder}, we can further reduce the vocabulary by ignoring words that occur less than a certain number of times and replacing them with "UNKNOWN".

After reducing the vocabulary, we can directly make token embedding, and the parameters of the embedding layer are updated during training. For tokens which can be split into several words, we average the embeddings of these words as their embeddings. Finally we get embeddings of all nodes and use them as the initial node representations.

\subsection{hyper-parameters and other configures}

We use SparseGGNN in Microsoft's dpu\_util library. \footnote{https://github.com/microsoft/dpu-utils/tree/master/python/dpu\_utils} The size of embedding vector is 64. We use GRU as the rnn cell of the network. The activation function of gru is RELU. The number of propagation steps is 8. There are four layers in the succeeding MLP. Their sizes are 64, 32, 16, 6. The activation function of the first three layer is RELU. 

\section{Evaluation}

\subsection{dataset}
We collected a dataset from 29 top-stared open-source java projects from Apache on Github. Then we filtered out the files with log sentences. After preprocessing, we got 8,000 labeled log samples. We selected the samples in 25 projects as seen data and divided them into train set, validation set and test set in the proportion 7-1-2 and use the samples in the other 4 projects as unseen data to examine the generalization capability of this model.
\subsection{evaluation metrics}
We use AUC and accuracy to evaluate this model. AUC is the area under ROC curve, and it is used to evaluate the degree of discrimination that is achieved by a model\cite{hengloglevel}. Higher AUC indicates better performance on discrimination of a model. 
\subsection{baseline}
We compare this model with the performance of random-guessing model, RNN model and Heng Li's model\cite{hengloglevel}. The random guessing model randomly assign labels to samples and doesn't learn from the train set. The RNN(recurrent neural network) model is commonly used to process serial inputs. It takes all the sentences in the code as input and learn the structural and semantic information of the code. Heng Li's team analyzed the structural information of log sentences, extracted the key features of log level from case study and used ordinal regression to determine which log level should be chosen. 
\subsection{quantitative evaluation}
Table 1 shows the performance of this model comparing to the baseline models.

\begin{table}[h!]  
  
  \centering  
  \fontsize{8}{10}\selectfont  
  \caption{performance of this model}  
  \label{tab:table1}  
    \begin{tabular}{|c|c|c|c|c|}  
    \hline  
    \multirow{2}{*}{Model} & \multicolumn{2}{c|}{Seen projects}&\multicolumn{2}{c|}{Unseen projects}\cr\cline{2-5}  
    &AUC&accuracy&AUC&accuracy\cr  
    \hline  
    this model& 0.880 & 0.657& 0.760 & 0.435\cr\hline  
    random guess& 0.5 & 0.167 & 0.5 & 0.167\cr\hline  
    LSTM& 0.7321& 0.627 & 0.728 & 0.457\cr\hline  
    Heng Li's model& 0.75-0.81 & - & - & -\cr\hline  
    this model+ RNN& 0.900 & 0.668& 0.796 & 0.460\cr\hline  
    \end{tabular}  
\end{table}  

On the 25 seen projects, this model beat rnn and Heng Li's model in AUC, with an accuracy of 65.7\%. On the 4 unseen projects, the AUC and accuracy are lower than seen projects. It is natural that this model and RNN model's performance is poor on unseen dataset, because logging standard varies in different projects. Although the performance of this model on unseen dataset is less decent, it still outperformed RNN and has an AUC of 0.760. We further combined this model with RNN through weighted addition of prediction arrays and achieved an AUC of 0.900 on seen projects.
\subsection{qualitative evaluation}
Figure 3 shows the confusion matrix of this model. The frequency of fatal in the 25 projects is less than 0.1\% so we omitted the label fatal in the matrix. \\
As is shown in the graph, the correct log level account for the highest proportion for all true log levels except trace, because the use of trace varies in different projects. Even if a log sentence is assigned to the wrong log levels, they are most likely to be assigned to the log levels adjacent to their true log levels, which means the errors tend to be acceptable.
\begin{figure}
  \includegraphics[width=\linewidth]{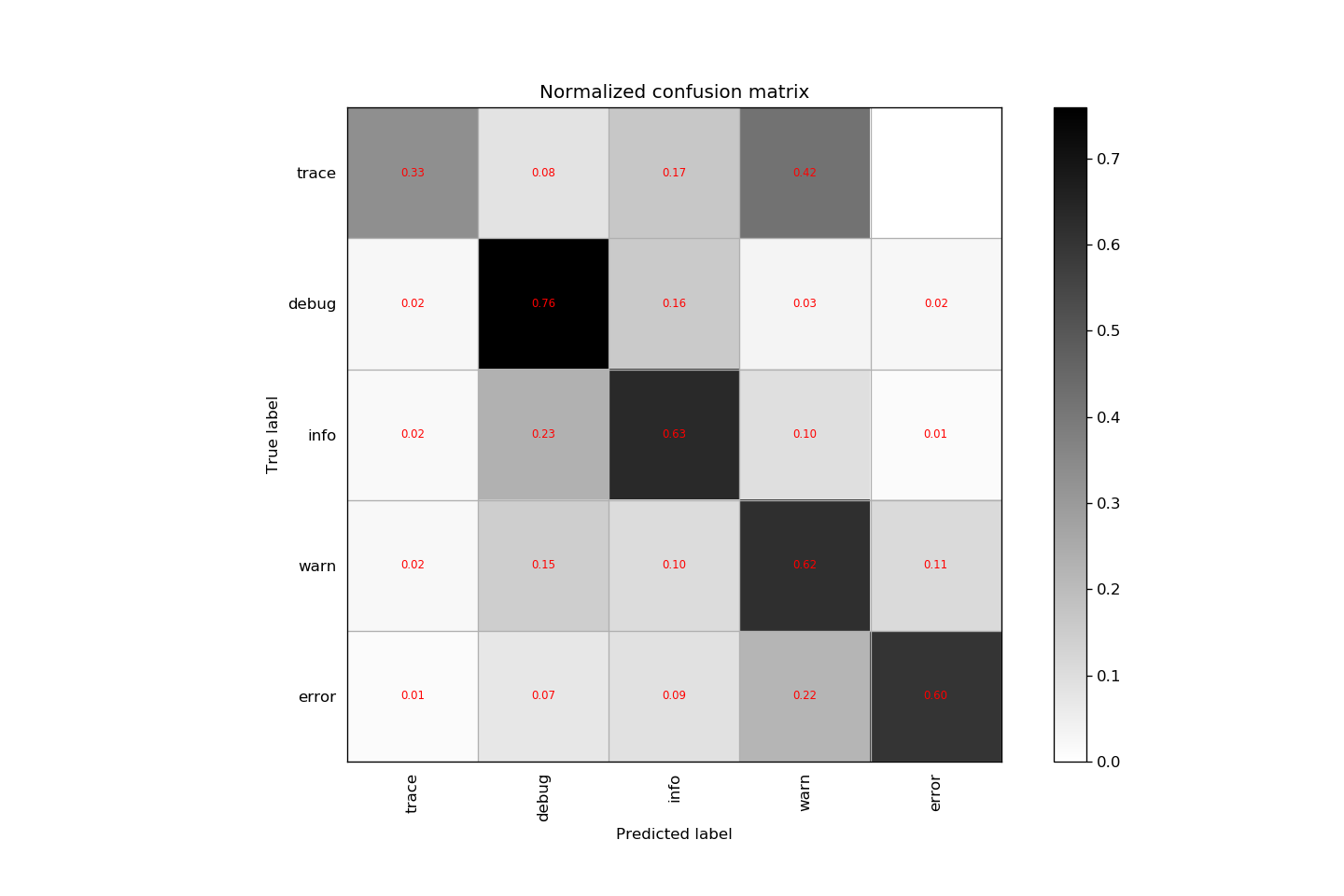}
  \caption{Confusion Matrix of this model}
  \label{fig:cofusion_matrix}
\end{figure}
\\
The code blocks below show examples of determining log levels. 
\begin{lstlisting}
if (t != null) {
      log.error(errMsg, t);
      throw new YarnException(errMsg, t);
    } else {
      log.error(errMsg);
      throw new YarnException(errMsg);
    }
\end{lstlisting}
After the preprocessing, log statement is replaced with semicolons, and the processed code is fed into this model. The log level of the log statements is to be predicted.
\begin{lstlisting}
if (t != null) {
      ;
      throw new YarnException(errMsg, t);
    } else {
      ;
      throw new YarnException(errMsg);
    }
\end{lstlisting}
The location of the log statements is connected to the statement 'throw' in the next line through NextToken edge, suggesting a high log level. Therefore, it is predicted as log level 'error' with high confidence and the prediction array is [0.055, 0.063, 0.084, 0.004, 0.794, 0.0002] for log levels [trace, debug, info, warn, error, fatal].
Another example is like
\begin{lstlisting}
  public long getCumulativeCpuTime() {
    if (LOG.isDebugEnabled()) {
      LOG.debug("CPU Comparison:" +
          procfs.getCumulativeCpuTime() + " " +
          cgroup.getCumulativeCpuTime());
    }
    return cgroup.getCumulativeCpuTime();
  }
\end{lstlisting}
After the preprocessing, the code block looks like
\begin{lstlisting}
  public long getCumulativeCpuTime() {
    if (LOG.isDebugEnabled()) {
    ;
    }
    return cgroup.getCumulativeCpuTime();
  }
\end{lstlisting}
From the statement 'LOG.isDebugEnabled()', the model knows the log statement is of low log level, but due to lack of information in the block, it could not determine the precise log level and gave a high degree of confidence to both 'debug' and 'info', and it is likely for the model to assign a slightly wrong log level to the statement and the prediction array is [0.025, 0.369, 0.595, 0.0003, 0.011, 1.9e-6] for log levels [trace, debug, info, warn, error, fatal], indicating high possibility for both debug and info.
\section{Discussion \& Conclusion}

Although many researchers treated code as a kind of natural language, code presents the properties of graphs in many ways. There are many complicated relationships in code and code is highly structured. So using graph neural network technology to process code is a good choice in some scenarios. Predicting log verbosity level is an appropriate example. In fact, predicting log verbosity level is a bit different from code recommendation. It's more like a code classification or summarization problem. Essentially, the task is to estimate the severity of the problem associated with a code block. If the problem is not serious, we are excepted to give a "info" or "warn". On the contrary, an "error" or "fatal" is given. This requires us to have a certain degree of certainty about the overall logic of the program which is beyond the reach of traditional NLP method. By using graph neural network, we can add custom edges to establish relationships between seemingly unrelated nodes so that the analysis process of the model can be guided and accelerated. As GNN can prove itself effective in this task, we are confident that it is also suitable for handling other similar tasks like defect detection and traceability. Furthermore, we can apply new research in the field of graph neural networks to create new models to deal with other programming problems. For example, graph kernels\cite{yanardag2015deep} may be applied to program classification. Attention mechanism can also be added to GNN models for programming problems. \cite{shang2018edge}In sum, applying GNN to code issues has a bright future.

\bibliographystyle{unsrt}  
\bibliography{references}

\end{document}